\documentstyle[12pt,epsf]{article}

\def\yzero{\smash{\hbox{$y\kern-4pt\raise1pt\hbox{${}^\circ$}$}}}

\def\-{\hphantom{-}}

\def\s2{\frac{1}{\sqrt2}}
\def\s22{\frac{1}{2\sqrt2}}

\def\beq{\begin{equation}}
\def\eeq{\end{equation}}
\def\beqa{\begin{eqnarray}}
\def\eeqa{\end{eqnarray}}
\newcommand{\bel}[1]{\begin{equation}\label{#1}}

\def\diag{{\rm diag \,}}
\def\Tr{{\rm Tr \,}}

\def\IF{\relax{\rm I\kern-.18em F}}
\def\II{\relax{\rm I\kern-.18em I}}
\def\IP{\relax{\rm I\kern-.18em P}}
\def\IR{\relax{\rm I\kern-.18em R}}

\def\IC{{\bf C}}

\def\NN{{\cal N}}

\def\IZ{{\bf Z}}
\def\cp#1{\relax\ifmmode {\IP\kern-2pt{}_{#1}}\else $\IP\kern-2pt{}_{#1}$\fi}
\def\e13{e^{2\pi i/3}}

\newcommand{\drawsquare}[2]{\hbox{%
\rule{#2pt}{#1pt}\hskip-#2pt
\rule{#1pt}{#2pt}\hskip-#1pt
\rule[#1pt]{#1pt}{#2pt}}\rule[#1pt]{#2pt}{#2pt}\hskip-#2pt
\rule{#2pt}{#1pt}}

\newcommand{\fund}{\raisebox{-.5pt}{\drawsquare{6.5}{0.4}}}
\newcommand{\antifund}{\overline{\fund}}

\begin{document}
\pagestyle{empty}
        
\makeatletter
\@addtoreset{equation}{section}
\makeatother
\renewcommand{\theequation}{\thesection.\arabic{equation}}







\rightline{IASSNS--HEP--98/88}
\rightline{\tt hep-th/9811004}
\vspace*{1cm}
\begin{center}
\LARGE{Brane Configurations for Branes at Conifolds}\\
\large{Angel M. Uranga}\\
{\em School of Natural Sciences, Institute for Advanced Study} \\
{\em Olden Lane, Princeton, NJ 08540}\\

\vspace*{2cm}

{\small {\bf Abstract}}
\end{center}
{\small
We study the T duality between a set of type IIB D3 branes at 
non-orbifold threefold singularities, and type IIA configurations of D4 
branes stretched between relatively rotated NS fivebranes. The 
four-dimensional 
$\NN=1$ field theories on the D3 brane world-volume can be easily 
described using the IIA brane configuration. These models include families 
of chiral theories continuously connected to the theories appearing in 
brane box models (or D3 branes at orbifold singularities). We propose that 
phase transitions in the K\"ahler moduli space of the  singularities are 
related to the crossing of rotated NS fivebranes in the T dual picture, 
and thus to Seiberg's duality in one of the gauge factors. We also 
comment on the inclusion of orientifold planes in the IIA brane picture.
}

\setcounter{page}{1}
\pagestyle{plain}
\renewcommand{\thefootnote}{\arabic{footnote}}
\setcounter{footnote}{0}
\newpage

\section{Introduction}

Configurations of NS fivebranes and D branes in string theory provide a 
useful and intuitive technique to study supersymmetric field theories in 
several dimensions (see \cite{review} for a review with extensive 
references). A particularly interesting case is that of  
four-dimensional $\NN=1$ gauge theories.  

There are several approaches in the construction of these models 
\footnote{Let us also mention the realizations in \cite{oovafa, lpt1}, 
also in this spirit.}. We briefly recall their basic ingredients. In 
\cite{hz}, type IIB configurations of NS branes (of two kinds) and D5 
branes were introduced to realize large families of chiral theories. 
These `brane box models'  have proved especially successful in the 
construction of $\NN=1$ finite field theories \cite{hsu}.

A different construction, pioneered in \cite{egk}, makes use of type IIA 
NS fivebranes (with relative rotations) and D4 branes stretching between 
them. These `rotated brane configurations' generically realize 
non-chiral theories, 
but have been particularly useful in deriving 
large classes of Seiberg dual pairs \cite{seiberg}, and in providing 
some exact results {\em via} their lifting to M-theory (see 
{\em e.g.} \cite{wittenn1,oz,biksy}). 

Finally, it is possible to realize gauge field theories on the worldvolume 
of D3 brane probes of a certain spacetime background. The simplest example 
is placing D3 branes on a $\IC^3/\Gamma$ singularity, with 
discrete $\Gamma\subset SU(3)$. The field theories, studied in \cite{dgm} 
along the lines of \cite{dm}, are generically chiral, and coincide with 
the 
theories obtained using certain brane box models. In \cite{hanur} it 
was shown that both constructions are related by a T duality, which 
transforms the D5 branes into D3 branes, and the NS fivebranes into the 
singular geometry. These relations between different pictures of the same 
theory are always illuminating. For instance, in the case at hand, the 
finiteness results of \cite{hsu} are closely related, using the 
singularity picture, to the conjectured AdS/CFT correspondence 
\cite{malda,gkp,witads}.

It is natural to seek the generalization to D3 brane probes in 
non-orbifold singularities. The simplest case, the conifold singularity, 
has been analyzed in \cite{kw}, where a long distance field theory with 
the appropriate symmetries and physical behaviour was proposed, as arising on 
the D3 brane world-volume.

Our purpose in the present paper is to explore the system of D3 branes at 
other non-orbifold singularities. The basic tool we exploit is to quotient 
the conifold variety $X$ by an appropriate discrete isometry group 
$\Gamma$. By 
determining the action of $\Gamma$ on the field theory, and keeping 
the dynamics of only the invariant states, the resulting gauge theory 
describes the D3 brane probes on the quotient $X/\Gamma$ (this type of 
`orbifolding' of the field theory has been mainly studied for D3 branes on 
flat space \cite{lnv}).  Thus we construct large classes of chiral 
and non-chiral $\NN=1$ field theories, with quartic superpotentials 
inherited from the conifold theory of \cite{kw}.

We also show that these geometries are related to type IIA brane 
configurations of rotated NS fivebranes and D4 branes by a T duality along 
a direction transverse to the NS branes. This transformation, studied in 
\cite{keshav} for the conifold, maps the D4 branes to D3 branes, and 
the NS fivebranes to the singular geometry. 
This relation nicely parallels the map between brane boxes and branes at 
orbifold singularities. We expect an interesting interplay of 
results from both pictures. Several results in this paper actually stem 
from T dual ways of looking at the same phenomenon: i) We argue that the 
continuation past infinite coupling in one of the gauge factors, 
realized as the crossing of two relatively rotated NS fivebranes in the 
type IIA brane picture, corresponds to transitions in the K\"ahler moduli 
space of the singular variety. ii) We describe explicitly the  field 
theories of D3 branes at certain singularities which are partial smoothings 
of quotients of the conifold. The smoothing is mapped to the removal of 
certain NS fivebrane in the type IIA configuration, and to following a 
certain baryonic Higgs branch in the field theory. iii) We can generalize 
the constructions by including orientifold planes in the IIA brane 
picture. This corresponds to performing an orientifold projection in the 
singular geometry, whose direct analysis would be difficult without the 
guide of the T dual brane model.

The paper is organized as follows. In Section~2 we derive the T duality 
between D3 branes at the conifold geometry and the type IIA configuration 
of rotated NS branes and D4 branes. In Section~3 we consider singularities 
obtained as quotients of the conifold, and also find the T dual IIA brane 
models. We describe the D3 brane world-volume field theories, and perform 
some consistency checks. We also show that K\"ahler transitions in the 
singularity picture correspond to crossings of rotated NS fivebranes. In 
Section~4 we study D3 branes on partial resolutions of these 
singularities, describing the corresponding IIA pictures, and the 
resulting field theories. In Section~5 we consider quotients yielding 
chiral gauge theories. These models have Higgs branches along which the 
field theories correspond to finite brane box models (equivalently, D3 
branes at $\IC^3/\Gamma$ singularities). In Section~6 we discuss the 
inclusion of orientifold planes. Section~7 contains some final comments.

After this work was completed, we noticed reference \cite{mp}, where a 
different approach to the problem of D3 branes at non-orbifold 
singularities is described. We expect further progress in the 
understanding of this system from the combination of different viewpoints.

\section{Brane configuration for the conifold}

Our starting point is a system of $N$ D3 branes sitting at the simplest 
non-orbifold threefold singularity, the 
conifold. The following analysis is similar to that of 
\cite{keshav}. The 
equation for the conifold
\beq
xy=zw
\label{conifold}
\eeq
allows us to interpret the variety as a $\IC^*$ fibration over the $\IC^2$ 
parametrized by $z$, $w$. That is, for generic values of $z$,$w$, the 
variables $x$, $y$ describe a $\IC^*$, since (\ref{conifold}) can be used 
to relate $x$ and $y$ (as long as they are not zero or infinite). The 
fiber degenerates to two intersecting complex planes when the right hand 
side of (\ref{conifold}) vanishes, {\em i.e.} along the (complex) curves 
$z=0$ and $w=0$ on the base.

Now we would like to perform a T duality along the $U(1)$ orbit in the 
$\IC^*$ fiber (the orbit is composed of points which are related by $x\to 
\lambda x$, $y\to \lambda^{-1} y$, with $\lambda$ a complex number of unit 
modulus) \footnote{To be more precise, we would like to perform the T 
duality in a global model where this $S^1$ has constant radius far away 
from the degenerations.}. Following standard arguments \cite{bsv} 
\footnote{Notice however that the T duality we are performing differs 
from that considered in \cite{bsv}}, the degenerations 
in the fibration denote the presence of NS fivebranes in the 
T dual picture. In our case, we find two NS branes, which 
span the directions, say, 012345 and 012389. We will denote them by NS and 
NS$'$ branes, respectively. The T dual space is otherwise flat, and one 
direction transverse to the NS fivebranes, say 6, is compact.

Notice that the two NS fivebranes need not be located at the same position 
in $x^6$. Their distance in $x^6$ is determined by the period of the NS 
two-form over the collapsed two-cycle in the conifold, whose value is not 
specified by the geometry (\ref{conifold}). So, generically the NS 
fivebranes do not intersect, and define two intervals in the circle 
parametrized by $x^6$.

\begin{figure}
\centering
\epsfxsize=3.5in   
\hspace*{0in}\vspace*{.2in}
\epsffile{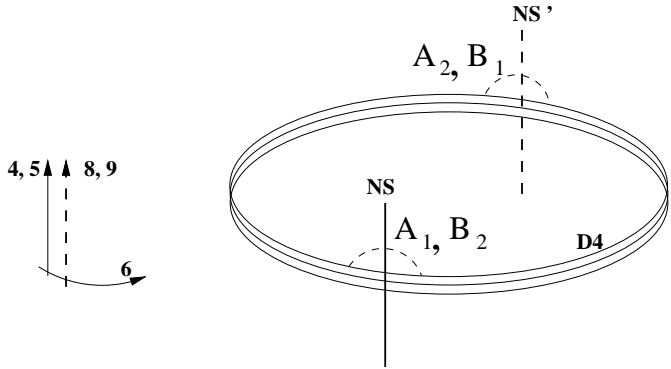}
\caption{\small The type IIA brane configuration T dual to the system of 
D3 branes at the conifold singularity. In this and the following pictures, 
we represent the NS brane as a continuous vertical line, and the NS$'$ 
brane as a discontinuous vertical line. Since they are oriented along 
different directions (45 vs. 89), the D4 branes cannot separate and there 
is no Coulomb branch. Nevertheless, for the sake of clarity, we have 
depicted the D4 branes as slightly separated. We also show the chiral 
fields arising from open strings (depicted as a discontinuous curve) 
stretching between the D4 branes.} 
\label{coni} 
\end{figure}

Finally, the D3 branes located at the conifold point in the singularity 
picture are mapped to D4 branes wrapping $x^6$. The resulting picture is 
shown in Figure~\ref{coni}.

Type IIA brane configurations involving D4, NS and NS$'$ branes have been 
considered in the literature (see \cite{review} for references) in the 
case of non-compact $x^6$. Those analyses allow to read off the spectrum 
of the four-dimensional $\NN=1$ field theory in our compact $x^6$ case as 
well.

The gauge group is $SU(N)^2 \times U(1)$ (throughout this paper we will 
often ignore these decoupled $U(1)$ factors), and there are chiral 
multiplets $A_1$, $A_2$ in the representation $(\fund,\antifund)$, and 
$B_1$, $B_2$ in the $(\antifund,\fund)$. The adjoint chiral multiplets 
$\Phi$, $\Phi'$ that would be massless if the NS fivebranes were parallel 
($\NN=2$ supersymmetric case \cite{witten4d}) receive a mass due to their 
relative rotation \cite{barbon}, which softly breaks the supersymmetry 
down to $\NN=1$. The superpotential is
\beqa
W= A_1 \Phi B_2 - A_2 \Phi B_1 - A_1 \Phi' B_2 + A_2 \Phi' B_1 
+ m \Phi^2 - m \Phi'^2.
\label{superpotential}
\eeqa
The adjoint masses are opposite since the two intervals have opposite 
rotation angle between their left and right NS fivebranes.

After integrating out the massive adjoints, the superpotential 
reads \footnote{Notice that the NS fivebranes are at right angles, so the 
adjoint mass is naively infinite. However, it is clear that the 
brane configuration corresponds to a field theory with a non-vanishing 
quartic interaction, since it must have a possible Higgssing to 
the $\NN=4$ theory, where there is a cubic superpotential.} 
\beq 
W \propto \Tr (A_1 B_1 A_2 B_2) - \Tr (A_1 
B_2 A_2 B_1) \propto \epsilon^{ij} \epsilon^{kl} \Tr (A_i B_k A_j B_l)
\label{superpkw}
\eeq
The field content mentioned above, and the superpotential (\ref{superpkw}) 
define the field theory proposed in \cite{kw} as arising on the 
world-volume of D3 branes at the conifold singularity. We have rederived 
the result by T dualizing the geometry to a more familiar brane 
configuration. Notice that the relation of the conifold theory to the 
softly broken $\NN=2$ theory was already uncovered in \cite{kw}.

The type IIA brane configuration has the disadvantage that the $SU(2)^2$ 
global symmetry of the field theory (under which the $A_i$ and $B_i$ 
fields transform in the $({\bf 2}, {\bf 1})$ and $({\bf 1}, {\bf 2})$ 
representations, respectively) is not manifest, whereas in the conifold 
picture they are realized geometrically \footnote{The $U(1)_R$ 
R-symmetry is realized geometrically in the IIA configuration as well.}. 
However, brane configurations of this type often provide interesting 
insights on many issues. For instance, the inclusion of additional 
flavours 
in the field theories is straightforward in the IIA brane configuration, 
whereas in the IIB T dual it corresponds to including D7 branes, which 
introduce unpleasant dilaton-axion variations. So we feel it is worth 
exploring other field theories that can be realized in the IIA framework.

In the following sections we study D3 branes on other singularities, in 
some cases related to the conifold by a quotient, and describe 
candidate T dual IIA brane configurations.

\section{Quotient of the conifold (I)}

In this section we describe a quotient of the conifold, and the field 
theory arising from D3 branes located at such singularity. We also present 
the T dual  brane configuration, and provide some consistency checks  
of our identification.

\subsection{Description}

The knowledge of the field theory on D3 branes at the conifold can be 
exploited to analyze other singularities which are not orbifolds of 
$\IC^3$. Simple examples of such worse 
singularities can be obtained by taking quotients of the conifold by 
some discrete symmetry. The conifold variety has an isometry group 
$SU(2)\times SU(2)\times U(1)_R$ (under which $x,y,z,w$ 
transform in the $({\bf 2}, {\bf 2})_{+1}$ representation). In order to 
preserve $\NN=1$ supersymmetry, the discrete groups will be embedded in 
the $SU(2)^2$ part, leaving the $U(1)_R$ R-symmetry untouched.

We will not attempt a general classification of such quotients, but 
present a few examples. A simple case, whose analysis we perform in the present 
section, is the $\IZ_k$ action generated by:
\beqa
x & \to & e^{2\pi i/k} x \nonumber \\
y & \to & e^{-2\pi i/k} y
\label{quot1}
\eeqa
leaving $z$ and $w$ invariant. By introducing the invariant variables 
$x'=x^k$, $y'=y^k$, $t=xy$, the final variety is described by the 
equations $x' y'=t^k$, $t=zw$, or equivalently by the expression
\beq
x' y'=z^k w^k
\label{singuk}
\eeq
This singularity, in the form $(z_1^2 +z_2^2)^k + z_3^2 + z_4^2=0$, has 
appeared in \cite{bsv,oovafabh}.

The field theory on the D3 branes can be obtained by an orbifolding 
procedure \cite{lnv} (see also \cite{schmaltz} for a general recipe in a 
purely field theoretical context). It 
is easy to see that the geometric action (\ref{quot1}) 
corresponds to the following action on the field theory chiral multiplets
\beqa
\theta: & A_1  \to & \theta A_1 \nonumber \\
        & A_2  \to & \theta^{-1} A_2 \nonumber \\
        & B_1  \to & \theta B_1 \\
        & B_2  \to & \theta^{-1} B_2 \nonumber 
\label{quot1ab}
\eeqa
with $\theta=e^{\pi i/k}$. 
To show this, recall the existence of a maximal Higgs branch where the 
matrices $A_1$, $A_2$, $B_1$, $B_2$ are diagonal and thus commuting. The 
$i^{th}$ eigenvalues $a_1$, $a_2$, $b_1$, $b_2$, parametrize the position 
of the $i^{th}$ D3 brane \footnote{The Higgs branch is obtained by taking 
all such possible distributions of branes and dividing by the permutation 
group.} on the conifold by the relation 

\beq
x=a_1 b_1 \quad ;\quad y=a_2 b_2 \quad ;\quad z= a_1 b_2 \quad ;\quad 
w=a_2 b_1.
\eeq 

The action (3.3) must be embedded in the gauge 
degrees of freedom, which amounts to choosing an action on the Chan-Paton 
factors of the D3 branes. Starting with a conifold theory with group 
$SU(M k)\times SU(M k)$, we choose these embeddings to be given by the 
matrices
\beqa
\gamma_{\theta}^{(1)} & = & \diag ({\bf 1}_M, \theta^2 {\bf 1}_M, \ldots, 
\theta^{2k-2} {\bf 1}_M) \nonumber \\
\gamma_{\theta}^{(2)} & = & \diag (\theta {\bf 1}_M, \theta^3 {\bf 1}_M, 
\ldots, \theta^{2k-1} {\bf 1}_M)
\label{cp1}
\eeqa
acting on the fundamental representations of the first and second factor, 
respectively. Here ${\bf 1}_M$ denotes the rank $M$ identity matrix. The 
generalization to other embeddings with different number of 
entries for different eigenvalues is straightforward. In this respect, we 
should mention that in the singularity picture it is not obvious whether 
all such embeddings are consistent, since consistency conditions manifest 
as cancellation of tadpoles and these are not easy to compute for spaces 
other than orbifolds of flat space. The T-dual brane picture we will 
construct below, however, shows explicitly that all choices are 
consistent, and correspond to putting different numbers of D4 branes in 
the different $x^6$ intervals (this is analogous to the $\NN=2$ case in 
\cite{karch}).

Going back to our choice of Chan-Paton factors (\ref{cp1}), it is a simple 
matter to perform the projection on the fields of the 
theory. Each of the factors in the original gauge group $SU(Mk)\times 
SU(Mk)'$ splits into $k$ identical $SU(M)$ factors, so the 
final $\NN=1$ vector multiplets form the group \footnote{As usual in 
four-dimensional theories, the $U(1)$ factors (save for the overall one) 
are not present in the low-energy theory \cite{witten4d}.} 
\beq
SU(M)^k \times {SU(M)'^k} \; =\; \prod_{i=1}^k SU(M)_i \times 
\prod_{j=1}^k SU(M)'_{j} 
\eeq
where we have introduced labels to distinguish the factors.

The different $\NN=1$ chiral multiplets suffer different projections 
depending on their global and gauge quantum numbers. For instance, the 
field $A_1$, in the $(\fund,\antifund)$ of the initial group, has a 
projection
\beq
\gamma_{\theta}^{(1)} A_1 (\gamma_{\theta}^{(2)})^{-1} =\theta A_1
\eeq
There are $k$ surviving chiral multiplets transforming in the 
representation $(\fund_{i+1},\antifund\; '_i)$ of $U(M)_{i+1} \times 
U(M)'_i$, for $i=1,\ldots,k$. We will denote these fields by 
$(A_1)_{i+1,i}$. 

Analogously, we obtain $k$ fields $(A_2)_{i,i}$ in the 
$(\fund_i,\antifund\; '_i)$, fields $(B_1)_{i,i}$ in the 
$(\fund\; '_i,\antifund_i)$, and fields $(B_2)_{i,i+1}$ in 
the $(\fund\; '_i,\antifund_{i+1})$ \footnote{Notice that for A-type 
fields the first (resp. second) index refers to unprimed (resp. primed) 
gauge factors, whereas for B-type fields the first (second) index refers 
to primed (unprimed) representation. This notations is convenient to 
suggest the appropriate contractions in the superpotential couplings.}. 
Notice that all these models are non-chiral.

The superpotential of the resulting theory is
\beq
W \propto \sum_{i=1}^k (A_1)_{i+1,i}\, (B_1)_{i,i}\, (A_2)_{i,i}\, 
(B_2)_{i,i+1} 
-\sum_{i=1}^k (A_1)_{i+1,i}\, (B_2)_{i,i+1}\, (A_2)_{i+1,i+1}\,  
(B_1)_{i+1,i+1}.
\label{superpot1}
\eeq

We can try to recover this field theory by a brane configuration related 
to the singularity via the T duality we discussed in Section~2. In order 
to do that, we interpret the defining equation $x' y' =z^k w^k$ as a 
$\IC^*$  fibration over the $\IC^2$ parametrized by $z$, $w$. The fiber 
degenerates again at the two complex planes $z=0$, and $w=0$. However, in 
this case the degeneration is worse, and the T dual picture contains $k$ 
NS branes and $k$ NS$'$ branes. After taking into account the 
possibility of having B-fields in the singularity picture, the positions 
of the NS fivebranes in the T dual compact coordinate $x^6$ may differ.
Again, the D3 branes sitting at the singularity become D4 branes wrapping 
$x^6$. The resulting IIA brane configuration looks like Figure~3 
\footnote{The orbifold 
field theory above is reproduced by a type IIA brane configuration where 
NS and NS$'$ branes are arranged in an alternating fashion. In 
subsection~3.2 we will consider the interpretation of other possible 
orderings}.

\begin{figure}
\centering
\epsfxsize=4in   
\hspace*{0in}\vspace*{.2in}
\epsffile{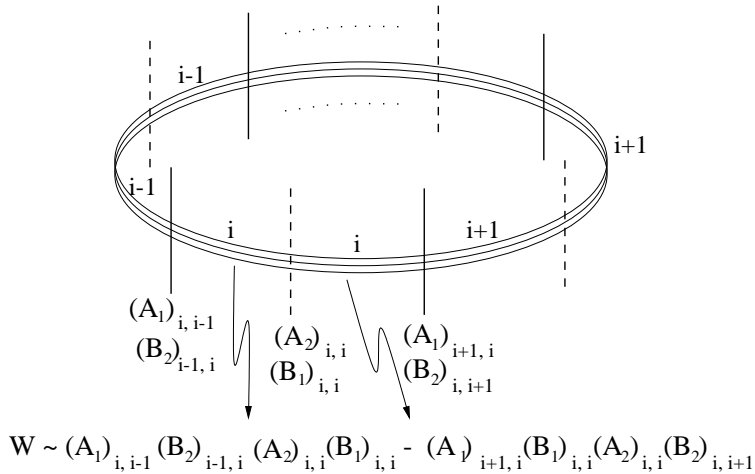}
\caption{\small Brane configuration T dual to a system of D3 branes at 
the $xy=z^k w^k$ singularity. The model consists in a set of $k$ NS and 
$k$ NS$'$ branes ordered along $x^6$ in an alternating fashion. The figure 
depicts the brane 
configuration in the vicinity of the $i^{th}$ pair of NS, NS$'$ branes.
We show the chiral fields in the bi-fundamental representations, and the 
superpotential quartic interactions mediated by the massive adjoints.
} 
\label{fig:orbi}
\end{figure}

The four-dimensional field theory on the D-branes can be obtained 
by noticing it corresponds to softly breaking a $\NN=2$ $SU(M)^{2k}$ model 
down to $\NN=1$ by appropriate adjoint masses. The gauge group is 
$\prod_{i=1}^k SU(M)_i$$\times \prod_{j=1}^k SU(M)'_j$, where unprimed 
factors correspond to intervals with a NS brane on their left end, and 
primed factors to intervals with NS$'$ branes on their left end.
It is easy to recognize the fields $(A_1)_{i,i-1}$, $(B_2)_{i,i-1}$
as arising from open strings stretching between the D4 branes 
associated to $SU(M)'_{i-1}$ 
and $SU(M)_i$; similarly, the fields $(A_2)_{i,i}$, $(B_1)_{i,i}$ appear 
from open strings joining the D4 branes in the intervals 
corresponding to $SU(M)_i$, $SU(M)'_i$.  Finally, the superpotential is 
obtained after introducing the adjoint masses and integrating these 
fields out. This type of exercise has been performed {\em e.g.} in 
\cite{hanbro,ahan}, and leads to the appearance of quartic 
superpotentials. The result in our case yields the superpotential 
(\ref{superpot1}). Thus the proposed IIA brane configuration reproduces 
the orbifold field theory.

\medskip

Before ending this subsection, we provide a few checks that support our 
identification. The first is the existence of mesonic Higgs branches in 
the field theory. These are  manifest in the IIA brane configuration, 
where they amount to splitting the D4 branes at two adjacent NS branes 
(of the 
same kind), and moving the pieces in the direction $45$, suspended 
between the NS branes. There are analogous  Higgs branches in which pieces 
of D4 branes travel along $89$, suspended between NS$'$ branes. An 
interesting hint is that the brane configuration localized near these 
traveling suspended D4 branes locally has $\NN=2$ supersymmetry. 

In the singularity picture, this phenomenon is accomplished by noticing 
that the singularity is not isolated. Actually, there is a curve of 
$\IC^2/\IZ_k$ singularities, parametrized by $w$, at $x=y=z=0$; and 
another similar curve, parametrized by $z$, at $x=y=w=0$. These curves 
arise from the set of points of the conifold which are fixed points under 
the $\IZ_k$ action (\ref{quot1}). The D3 branes at the origin can split 
into fractional branes \cite{douglas} which can travel along these curves of 
singularities, but not away from them. This description of Higgs branches 
is identical to that in \cite{karch} for two-fold singularities and 
\cite{hanur} for three-fold singularities. Along these branches, we have 
a set of D3 branes at a $\IZ_k$ ALE singularity, a system with $\NN=2$ 
supersymmetry.

Besides these mesonic branches, there are baryonic Higgs branches which, 
in the brane configuration, are realized as the removal of, say, one NS$'$ 
brane along $x^7$. In the field theory, it arises by giving diagonal vevs 
to the 
fields $(A_2)_{i,i}$, $(B_1)_{i,i}$, which triggers the breaking 
$SU(M)_i\times SU(M)'_i$$\to SU(M)_{i,\diag}$. Notice that one of the 
chiral multiplets is swallowed by the Higgs mechanism, whereas the other 
remains in the theory as a chiral multiplet in the adjoint. This 
remaining field parametrizes the 
possibility of moving the D4 branes along the two neighbouring NS branes.
One such brane configuration is depicted in Figure~\ref{fig:spp}. 
We will 
encounter again this type of brane configuration in Section~4, where we 
will see they are associated to partial resolutions of the singularity 
$xy=z^k w^k$.

These brane configurations can also be generalized by introducing D6 
branes. This amounts to adding fundamental flavours in the field theory. 
In the T dual picture, this corresponds to the addition of D7 branes, and 
so the configuration is better described in terms of F-theory. We will 
not pursue these very interesting generalizations in the present paper.

\subsection{Counting of marginal parameters}

In this subsection we are going to compute the number of exactly marginal 
operators in these field theories. Our motivation is to argue that the 
superpotential of the field theory we are studying can be defined, in 
analogy with \cite{kw}, as a 
marginal deformation around a conformal point of the free theory. Also we 
show that most of the marginal couplings have a clear interpretation in 
the IIA brane construction.

From the field theory point of view, the number of marginal couplings can 
be determined using the techniques in \cite{matt}, which were already 
exploited in \cite{hsu} for a similar counting in brane box models 
(equivalently, D3 branes at orbifolds of $\IC^3$). For notational 
clarity, let us momentarily denote the fields $(A_1)_{i+1,i}$, 
$(B_2)_{i,i+1}$, $(A_2)_{i,i}$ and $(B_1)_{i,i}$, by ${\tilde F}_i$, 
$F_i$, ${\tilde G}_i$ and $G_i$, respectively. The superpotential of the 
theory reads
\beq
W= \sum_{i=1}^k \lambda^{(1)}_i F_i {\tilde F}_i G_{i} {\tilde G}_{i} 
+ \sum_{i=1}^k \lambda^{(2)}_i {\tilde F}_i F_i {\tilde G}_{i+1} G_{i+1} 
\eeq
where we have allowed for arbitrary superpotential couplings. The 
parameter  space in the model is spanned by $2k$ gauge couplings and $2k$ 
superpotential couplings. 

The conditions for a conformal theory can be found by using the exact beta 
functions for these parameters \cite{sv1,sv2}. For the gauge couplings of 
$SU(M)_i$, $SU(M)'_i$, the vanishing of the beta function reads
\beqa
\beta_{g_i} & = & 2+\gamma_{F_{i-1}}+\gamma_{{\tilde F}_{i-1}} + 
\gamma_{G_{i}} +\gamma_{{\tilde G}_{i}} =0 \nonumber\\
\beta_{g'_i} & = & 2+\gamma_{F_i}+\gamma_{{\tilde F}_i} + \gamma_{G_{i}} 
+\gamma_{{\tilde G}_{i}} =0 
\label{betas}
\eeqa
where $\gamma_{X}$ is the anomalous dimension of the field $X$. For the 
superpotential couplings $\lambda^{(2)}_{i-1}$, $\lambda^{(1)}_i$, the 
vanishing of the beta functions are also given by the two equations 
(\ref{betas}), and do not provide independent constraints. 
Furthermore, there is one linear relation among the $2k$ 
conditions (\ref{betas}), $\sum_i \beta_{g_i}=\sum_i \beta_{g'_i}$. So we 
have a total of $2k-1$ conditions on $4k$ couplings, 
leading to a $(2k+1)$-dimensional manifold of RG fixed points on the 
parameter space \footnote{The conifold case is special in that the two 
superpotential couplings are equal due to a global 
$SU(2)\times SU(2)$ symmetry. This results in two marginal couplings 
\cite{kw} instead of the three our general counting indicates.}.

The existence of these marginal couplings is directly inherited from the 
marginal superpotential in the conifold theory. Thus we can use their 
existence to define our field theory in the infrared in analogy with the 
argument in \cite{kw}. In the absence of superpotential, the field theory 
with the proposed matter content flows to a conformal theory. There it is 
possible to turn on the marginal couplings, and we recover the theory of 
interest.

The type IIA brane configuration provides a geometric realization
of these marginal couplings. They correspond to the $k-1$ independent 
$x^6$ distances between NS branes of the same kind (adequately 
complexified by the shift in the type IIA RR $U(1)$ gauge field, or 
equivalently the positions of the NS branes in the eleventh 
dimension of M-theory \cite{witten4d}), the $k-1$ independent distances 
between NS$'$ branes, and the length of the $x^6$ coordinate 
(complexified to the complex structure of the torus parametrized by 
$x^6$, $x^{10}$ in M-theory). Another parameter corresponds to the 
relative positions of the sets of NS and NS$'$ branes.
Notice that a last marginal couplings seems to be not explicit in the 
brane construction.

\subsection{K\"ahler transitions vs. ordering of branes}

The proposal of a type IIA brane configuration T dual to the D3 branes at 
the singularity (\ref{singuk}) poses a puzzle. There exist many different 
brane configurations containing $k$ NS branes and $k'$ NS$'$ branes, 
which 
differ in the ordering of the fivebranes along the coordinate $x^6$. Even 
though these are physically different, they all can be claimed, by the 
arguments in section~3.1, to be T dual to a set of D3 branes at the space 
(\ref{singuk}). This singularity, however, does not seem to contain any 
degree of freedom corresponding to the multiple choices in the type IIA 
picture.
The resolution of the puzzle consists precisely in a proper identification 
of these choices in the singularity picture. We will argue that the 
different orderings of branes correspond to different phases in 
the K\"ahler moduli space of the singularity \footnote{I acknowledge R. von 
Unge for useful conversations on the contents of this section and on his 
work \cite{unge}. His comments encouraged me to correct several confusing 
statements present in a previous version.}.

The singular variety (\ref{singuk}) can be thought of as a smooth manifold in 
the limit in which a set of $\IP_1$'s shrinks to zero size. However, there are 
topologically different (but birrationally equivalent) smooth manifolds 
which can degenerate to the variety (\ref{singuk}). These manifolds are 
related by flop transitions (see {\em e.g.} \cite{agm,witphase}). Thus 
there are different ways of resolving the singularity by restoring the 
finite size of different sets of shrunk $\IP_1$'s. In string theory the 
size of these cycles is complexified by the corresponding period (B-field) 
of the type IIB NS two-form. The moduli space spanned by these `complexified 
sizes' is known as `complexified K\"ahler moduli space'.

Singularities in this moduli space arise at points where the size and 
B-field of a given cycle vanish. These loci are of complex codimension one 
and thus, paths interpolating between two points in moduli space can 
always avoid them. Even though this implies there are no true phase 
transitions in these interpolations, we will loosely use the term 
`K\"ahler phase transition' to denote  paths in moduli space that 
actually cross the singular point.

Our proposal is that for each possible configuration of NS and NS$'$ branes 
on the $x^6$ circle there is a phase in the complexified K\"ahler moduli 
space of the T dual singularity. Also, the exchange of adjacent NS and NS$'$ 
branes corresponds to K\"ahler transitions in which the B-field of a 
vanishing cycle changes sign. This can be understood as follows. Using the 
T duality map, it is easy to realize that when all NS and NS$'$ branes are 
located at $x^7=0$ all the two-cycles in the dual variety have vanishing 
real size. However, the B-fields on these cycles generically do not 
vanish, and they encode the positions of the type IIA NS fivebranes 
along the $x^6$ circle. The process of crossing a NS and a NS$'$ brane by 
moving them in $x^6$, while keeping them at $x^7=0$, is mapped to a 
transition where the B-field of the T dual two-cycle varies continuosly 
and changes sign in the process, while the real size of the cycle remains 
zero. The singular point in moduli space, where the 
two-cycle has zero size and B-field, corresponds to the NS and NS$'$ branes 
exactly intersecting. This point can be avoided by a slight 
$x^7$-separation of the NS and NS$'$ branes when the coincide in $x^6$, in 
the same way as the singularity in K\"ahler moduli space can be avoided 
by turning on a non-zero real size for the two-cycle when the B-field 
vanishes.

The natural context to study the geometry of K\"ahler moduli space is 
toric geometry. A detailed presentation of toric geometry is outside the 
scope of the present paper, so we will merely depict the relevant toric 
diagrams for convenience of the readers familiar with this formalism (we 
refer to \cite{fulton}, \cite{agm} for further details). Since the toric 
description is not essential for other sections, other readers are 
adviced not to worry about these alternative pictures. 

It would be nice to have a precise match between the different orderings 
of fivebranes in $x^6$ and the different configurations of B-fields. This 
would require a precise description of the B-field moduli space, but the 
details of this characterization in toric geometry are unknown to us. On 
the other hand, toric geometry provides a simple description of the moduli 
space of real sizes of the two-cycles, where the problem reduces
to determining the  different triangulations of a polygon associated to 
the singularity. Thus each ordering of the fivebranes in the non-compact 
direction $x^7$ corresponds to a particular triangulation of the polygon.

We may expect an analogously simple structure for the moduli space of 
B-fields. Notice that it should correspond to the ordering of the 
fivebranes in the {\em compact} direction $x^6$, so the global 
considerations should be different. In what follows we describe a 
suggestive correspondence between the different orderings of fivebranes
in $x^6$ with the different triangulations of a polygon, once we take into 
account certain `compactification prescriptions' for the diagram, to be 
described below. Notice that in these figures the triangulations do {\em 
not} represent small resolutions of the singularity (all two-cycles have 
vanishing size when all fivebranes sit at $x^7=0$), but somehow represent 
different configurations of B-fields on the two-cycles. The precise 
meaning of the triangulations in this context is thus somewhat unclear. 

\begin{figure}
\centering
\epsfxsize=4.5in   
\hspace*{0in}\vspace*{.2in}
\epsffile{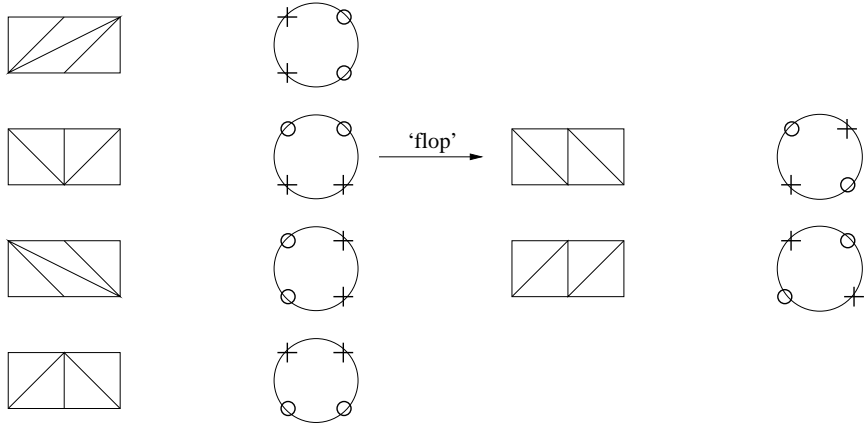}
\caption{\small Comparison between the possible orderings of two NS branes 
and two NS$'$ branes along $x^6$ and the triangulations of the diagram for 
the singularity $xy=z^2 w^2$. The NS branes are represented by crosses, and 
the NS$'$ branes as circles, sitting at points in a circle representing 
the $x^6$ direction. We also show one example of a transition, and 
the corresponding exchange of NS and NS$'$ branes in the T dual brane 
picture.} 
\label{flop}
\end{figure}

In Figure~\ref{flop} we have depicted the different triangulations of the 
diagram for the singularity $xy=z^2 w^2$, along with the corresponding brane 
configuration \footnote{Other examples can be worked out similarly, the 
basic rules  (for which I only have `empirical' evidence) being as 
follows: {\bf i)} Different toric diagrams related by a `rotation' of 
their $\IP_1$'s as explained below correspond to the same brane 
configuration (up to an irrelevant rotation along $x^6$); {\bf ii)} The 
existence, in the toric diagram, of triangles forming the diagram of 
$\IC\times(\IC^2/\IZ_k)$ corresponds to the existence of $k$ parallel 
adjacent fivebranes, forming a $\NN=2$ subsector.}.
 
\begin{figure}
\centering
\epsfxsize=4.5in   
\hspace*{0in}\vspace*{.2in}
\epsffile{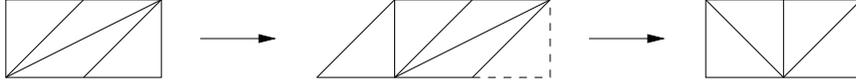}
\caption{\small Intuitive argument showing that two triangulations 
corresponding to identical brane configurations differ by a `rotation' 
among their $\IP_1$'s. In the first step, we `cut' the lower left triangle 
piece of the toric diagram and `glue' it on the other side. This amounts 
to making compact the previously non-compact $\IP_1$, and making 
non-compact a previously compact one. The second step is a mere re-drawing 
of the same toric diagram.} 
\label{rotation} 
\end{figure}

A subtlety in this identification is that some distributions of fivebranes 
only differ in a translation along $x^6$. The corresponding diagrams, 
on the other hand, look rather different. The explanation for this  lack of 
symmetry is that the variety contains a non-compact $\IP_1$, which can be 
understood in the toric diagram as arising from gluing together the two 
complex planes represented by vertical segments on the sides. This last 
$\IP_1$ `closes up' a chain of $\IP_1$'s (in analogy 
with the way the affine node of an extended $A_n$ Dynkin diagram closes up 
the line of nodes of the non-extended diagram). The translation in $x^6$ in 
the brane picture somehow transforms the `affine' $\IP_1$ into a regular one, 
and a regular $\IP_1$ into the new `affine' one. One such process is 
illustrated in Figure~\ref{rotation}, which should also clarify the 
systematics of our correspondence in Figure~\ref{flop}. This figure also 
illustrates the `compactification prescription' that we mentioned above.

It is nice to observe in the examples depicted in Figure~\ref{flop} how 
the K\"ahler transitions correspond to exchanges of pairs NS-NS$'$. However, 
some exchanges of branes relate toric diagrams which look very different. 
Again, this is the case when the additional `affine' $\IP_1$ is the 
`flopped' one.

Let us turn to the field theory interpretation of this transitions. 
From the analysis in \cite{egk}, the exchange of NS and NS$'$ branes is 
usually interpreted as $\NN=1$ duality \cite{seiberg} between the gauge 
theories described by the initial and 
final brane configurations. In our case, the gauge group contains many factors 
and the crossing should correspond to dualizing just one of them. Isolating 
this sector of the model, the `electric' theory belongs to a family of 
theories with gauge group $SU(N_c)$, $n$ flavours 
$Q,{\tilde Q}$, and $m$ flavours $Q', {\tilde Q'}$ all in the fundamental 
representation, 
and a quartic superpotential $W= (Q {\tilde Q})(Q' {\tilde Q'})$. The 
`magnetic' theory has group $SU(N_c-n-m)$, $n$ flavours $q, {\tilde q}$ 
and $m$ flavours $q',{\tilde q'}$, all in the fundamental, $n^2-1$ 
singlets mesons $M$, and $m^2-1$ singlet mesons $M'$. The superpotential 
is $W=qM{\tilde q}+ q'M'{\tilde q'}+ (q{\tilde q})(q'{\tilde q'})$. The dual 
pair can be obtained from the model in \cite{seiberg} upon deforming it 
with the quartic superpotential, and has also been derived using 
brane configurations in \cite{ahan}. In our type IIA construction, the 
transition of crossing the NS and NS$'$ branes reproduces this family of 
dual pairs. Different $m$, $n$, $N_c$ can be achieved by using 
different numbers of D4 branes on different intervals. In the singularity 
picture, they are mapped to fractional branes \cite{karch}. Observe that 
in our models the global symmetries of the field theory are gauged.

Notice that two theories related by brane crossing transitions form a dual pair
if the field theories flow to strong coupling in the infrared. As 
intuitively argued in \cite{oovafa}, in such case the NS fivebranes tend 
to come together and the two theories flow to the same configuration. 
According to this interpretation, the case $N_c=n=m$ in our particular 
example is quite subtle, since the corresponding field theory  has a 
marginal coupling. If it indeed corresponds to the distance between the 
fivebranes in the brane construction, there is no apparent reason why the 
transition gives a dual pair, since in the IR this distance would remain 
non-zero. It would interesting to gain some insight on this issue.

\section{Partial resolutions of the quotient of the conifold}

In this section our purpose is to analyze D3 branes on hypersurface 
singularities on $\IC^4$ of the form
\beq
xy=z^n w^m
\label{spp}
\eeq
In general, these singularities  are not quotients of the conifold, and 
thus the field theory on the D3 branes are not so straightforward to 
obtain. However, our T duality map still applies in analogy with previous 
sections, and one can use it to read off the field theory from the type 
IIA T dual brane configuration.

Following the usual argument, the T dual configuration contains $n$ NS 
branes and $m$ NS$'$ branes, located at certain values in the $x^6$ 
coordinate. As we know, the specific ordering possibilities are in 
one-to-one correspondence with the different choices of B-fields of the 
singularity. Given one such type IIA brane configuration, the 
field theory is easily determined. The gauge group is $U(M)^{n+m}$. There 
are bifundamental fields in the $(\fund,\antifund)$$+(\antifund,\fund)$ of 
adjacent gauge factors. Furthermore, there is an adjoint chiral multiplet 
whenever two NS fivebranes of the same kind are adjacent. 

The superpotential can be determined by starting with an $\IZ_k$ orbifold 
model as those studied in Section~3, with $k={\rm max} (m,n)$, and 
removing the adequate number of NS or NS$'$ branes. In field theory 
language, this corresponds to going into the appropriate baryonic Higgs 
branches. The outcome of this exercise can be summarized in the following 
rules
\begin{itemize}
\item Whenever two relatively rotated NS fivebranes are adjacent, there is 
a quartic superpotential for the fields living at the ends (the interaction is 
mediated by the massive adjoint): $W= \pm F{\tilde F} G {\tilde G}$. 
The sign is taken positive (negative) if a NS (NS$'$) brane lies at the 
left end of the interval.  
\item if two parallel NS fivebranes are adjacent, the superpotential is 
the usual $\NN=2$ coupling between  the adjoint and the chiral 
fields at the ends of the interval: $W= F\Phi {\tilde F}- G \Phi 
{\tilde G}$.
\end{itemize}
Notice that the superpotential (\ref{superpot1}) can be obtained from the 
brane configuration Figure~\ref{fig:orbi} by applying these rules.

Just to provide an explicit example, we describe the specific case of the 
singularity $xy=zw^2$. In order to reach that model, we start with a 
$\IZ_2$ quotient of the conifold, $xy=z^2w^2$, with group $SU(M)^4$, and 
matter fields 
{\small
\begin{center}
\begin{tabular}{ccccc}
    & $SU(M)_1$ & $SU(M)_2$ & $SU(M)_3$ & $SU(M)_4$ \\
$X_{12}$ & $\fund$ & $\antifund$ & $1$ & $1$ \\
$X_{21}$ & $\antifund$ & $\fund$ & $1$ & $1$ \\
$X_{23}$ & $1$ & $\fund$ & $\antifund$ & $1$ \\
$X_{32}$ & $1$ & $\antifund$ & $\fund$ & $1$ \\
$X_{34}$ & $1$ & $1$ & $\fund$ & $\antifund$ \\
$X_{43}$ & $1$ & $1$ & $\antifund$ & $\fund$ \\
$X_{41}$ & $\antifund$ & $1$ & $1$ & $\fund$ \\
$X_{14}$ & $\fund$ & $1$ & $1$ & $\antifund$ \\
\end{tabular}
\end{center}
}
The superpotential is
\beq
W \; = \; - X_{21}X_{12}X_{23}X_{32} + X_{32}X_{23}X_{34}X_{43} - 
X_{43}X_{34}X_{41}X_{14} + X_{14}X_{41}X_{12}X_{21}
\label{superpotz2}
\eeq

The removal of one NS brane corresponds to giving a diagonal vev to the field 
$X_{34}$. This breaks the gauge factors 3 and 4 to the diagonal 
combination, denoted 3 in the following. Thus the gauge group is 
$SU(M)_1\times SU(M)_2\times SU(M)_3$. The matter content is
\begin{center}
\begin{tabular}{cccc}
    & $SU(M)_1$ & $SU(M)_2$ & $SU(M)_3$ \\
$X_{12}$ & $\fund$ & $\antifund$ & $1$ \\
$X_{21}$ & $\antifund$ & $\fund$ & $1$ \\
$X_{23}$ & $1$ & $\fund$ & $\antifund$ \\
$X_{32}$ & $1$ & $\antifund$ & $\fund$ \\
$X_{31}$ & $\antifund$ & $1$ & $\fund$ \\
$X_{13}$ & $\fund$ & $1$ & $\antifund$ \\
$\Phi_3$ & $1$ & $1$ & {\rm Adj.} \\
\end{tabular}
\end{center}
where the field $\Phi_3$ is the former $X_{34}$ that transforms in the 
adjoint of the diagonal group.

\begin{figure}
\centering
\epsfxsize=2.5in   
\hspace*{0in}\vspace*{.2in}
\epsffile{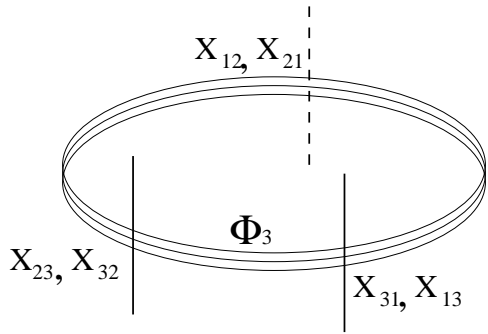}
\caption{\small Brane configuration T dual to a system of D3 branes at 
the $xy=zw^2$ singularity.}
\label{fig:spp}
\end{figure}

The brane configuration is depicted in Figure~\ref{fig:spp}. The 
superpotential 
that follows from (\ref{superpotz2}) reads
\beq
W = - X_{21}X_{12}X_{23}X_{32} + X_{23}\Phi_3 X_{32} - X_{13}\Phi_3 
X_{31} + X_{13}X_{31}X_{12}X_{21}.
\label{superpotspp}
\eeq
We can easily check this also follows from our above rules. Other examples 
can be worked out analogously.

The theory of D3 branes at the singularity $xy=zw^2$ was also studied in 
\cite{mp}. Even though our derivation is different, the field theory we 
have proposed agrees with that in \cite{mp}.

Again, it is a nice exercise to match the Higgs branches of the field 
theory using both string theory descriptions: the IIA brane configuration, 
and the branes at singularities. For instance, as mentioned at the end of 
subsection~3.2, baryonic Higgs branches are 
realized by the removal of fivebranes in the brane configuration. It is 
clear that this allows to relate different theories in this class, so let 
us discuss how our rules to determine the superpotential take that into 
account. 

Giving a diagonal vev to the field $X_{23}$ corresponds to removing one of 
the NS branes in the type IIA picture. The remaining configuration was 
proposed in section~2 as the T-dual to the conifold theory. This is neatly 
reproduced in the field theory, since the vev $<X_{23}>=v$ gives a mass to 
the $\Phi_3$, $X_{32}$. We integrate them out by using their equations of 
motion
\beq
X_{32} \; =\; {1\over v} {\tilde X}_{21} {\tilde X}_{12} \quad \;\quad
\Phi_3 \; =\; X_{21}X_{12}
\eeq
Here ${\tilde X}_{12}$, ${\tilde X}_{21}$ are the former $X_{13}$, 
$X_{31}$. The final superpotential reads
\beq
W= - X_{21}X_{12}{\tilde X}_{21}{\tilde X}_{12} + X_{21}{\tilde 
X}_{12}{\tilde X}_{21} X_{12},
\eeq
as corresponds to the conifold.

Instead, we can start with Figure~\ref{fig:spp}, and follow the baryonic 
branch corresponding to removing the NS$'$ brane. This is accomplished
by a diagonal vev for $X_{12}$. In this case no fields become massive, and 
the interactions (\ref{superpotspp}) become the superpotential of an 
$\NN=2$ $SU(M)^2$ elliptic model.

\medskip

\begin{figure}
\centering
\epsfxsize=3.5in   
\hspace*{0in}\vspace*{.2in}
\epsffile{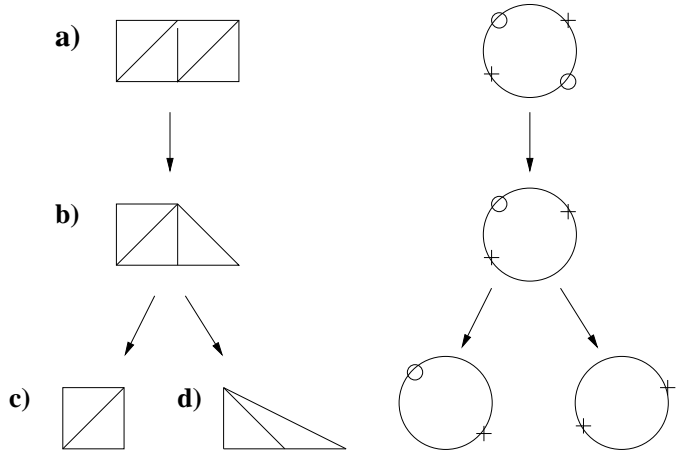}
\caption{\small The toric diagrams corresponding to the singularities a) 
$xy=z^2w^2$, b) $xy=zw^2$, c) $xy=zw$, d) $xy=z^2$. We have also depicted 
the corresponding T dual IIA brane configurations, illustrating that the 
removal of NS fivebranes corresponds to the partial resolution of the 
singularity.} 
\label{toric}
\end{figure}

The removal of the fivebranes is clearly mapped, in the picture of branes at 
singularities, to the partial resolution of the singularity. This can be 
seen in the equation for the variety, or very pictorically, in the toric 
diagrams of these spaces. In Figure~\ref{toric} we have depicted the toric 
diagrams of the singularities and how they are connected by the 
mentioned Higgs breakings.

\section{Quotients of the conifold (II): Chiral theories}

\subsection{$\IZ_k\times \IZ_l$ orbifolds}

Even worse singularities can be analyzed by performing a further quotient 
to the theories studies in Section~3. For instance, we can consider 
quotients of the conifold by a $\IZ_k\times \IZ_{l}$ action, whose 
generators act as as
\beqa
\alpha: & x \to & e^{2\pi i/k} \; x  \nonumber \\
       & y \to & e^{-2\pi i/k} y \quad\quad z,w \; \;{\rm invariant} \\
\beta: & z \to &  e^{2\pi i/l}\; z  \nonumber \\
        & w \to & e^{-2\pi i/l} w \quad\quad x,y \; \; {\rm invariant.} 
\label{quot2}
\eeqa
It is straightforward to define invariant variables and check that 
the resulting manifold can be represented as the hypersurface
\beq
x'^l y'^l \; =\; z'^k w'^k
\label{singukl}
\eeq
in $\IC^4$. This singularity has appeared in \cite{bsv}, in the form 
$(z_1^2+z_2^2)^k+ (z_3^2+z_4^2)^l=0$.

The action (\ref{quot2}) has its counterpart on the fields $A_i$, $B_i$ 
of the conifold 
theory:
\begin{center}
\begin{tabular}{lllllll}
$\theta$: & $A_1\to$ & $\theta\; A_1$ & & $\omega:$ & $A_1\to$ & $\omega\; 
A_1$ \\
          & $A_2\to$ & $\theta^{-1} A_2$ & & & $A_2\to$ & $\omega^{-1} A_2$ \\
          & $B_1\to$ & $\theta\; B_1$ & &  & $B_1\to$ & $\omega^{-1} B_1$ \\
          & $B_2\to$ & $\theta^{-1} B_2$ & &  & $B_2\to$ & $\omega\; B_2$ 
\\
\end{tabular}
\end{center}       
where $\theta=e^{\pi i/k}$, $\omega=e^{\pi i/l}$.

Starting with a conifold theory with group $SU(Mkl)\times SU(Mkl)$, the 
embedding on the gauge degrees of freedom can be realized by the 
following Chan-Paton matrices:
{\small
\beqa
\gamma_{\theta}^{(1)} & = & \diag ( 
{\bf 1}_M,{\bf 1}_M\ldots,{\bf 1}_M;\theta^2 {\bf 1}_M,\theta^2 
{\bf 1}_M\ldots,\theta^2 {\bf 1}_M; \ldots\ldots; \theta^{2k-2}{\bf 1}_M, 
\theta^{2k-2} {\bf 1}_M \ldots,\theta^{2k-2} {\bf 1}_M ) \nonumber \\
\gamma_{\theta}^{(2)} & = & \diag ( 
{\underbrace {\theta {\bf 1}_M,\theta {\bf 1}_M\ldots,\theta {\bf 
1}_M}_l};{\underbrace  {\theta^3 {\bf 1}_M,\theta^3 {\bf 1}_M 
\ldots, \theta^3 {\bf 1}_M}_l} ; \ldots\ldots; {\underbrace {\theta^{2k-1} 
{\bf 1}_M, \theta^{2k-1} {\bf 1}_M \ldots, \theta^{2k-1} {\bf 1}_M}_l} ) 
\nonumber \\
\gamma_{\omega}^{(1)} & = & \diag ( {\bf 1}_M, \omega^2 {\bf 1}_M 
\ldots, \omega^{2l-2} {\bf 1}_M; {\bf 1}_M, \omega^2 {\bf 1}_M 
\ldots, \omega^{2l-2}{\bf 1}_M; \ldots\ldots ;{\bf 1}_M, \omega^2 {\bf 1}_M 
\ldots,\omega^{2l-2} {\bf 1}_M ) \nonumber \\
\gamma_{\omega}^{(2)} & = & \diag ( 
\omega {\bf 1}_M, \omega^3 {\bf 1}_M \ldots, \omega^{2l-1} {\bf 1}_M; \omega 
{\bf 1}_M,\omega^3 {\bf 1}_M \ldots, \omega^{2l-1} {\bf 1}_M; 
\ldots\ldots; \omega {\bf 1}_M, \omega^3 {\bf 1}_M\ldots,\omega^{2l-1}
{\bf 1}_M
) \nonumber 
\eeqa
}

We stress that, contrary to the case in section~3.1, other choices of 
Chan-Paton factors may yield inconsistent theories. This is manifest since 
in this case the D3 brane world-volume theories are chiral and a generic 
choice of Chan-Paton matrices results in gauge anomalies in the field 
theory. The problem of determining the string consistency conditions in 
non-orbifold singularities seems rather complicated, and in the present 
work we will restrict to exploring the consistent choice shown above.
 
Using the projections induced by these actions, one can work out the field 
theory content. As expected, the gauge group in the orbifold models is
\beq
\prod_{i=1}^k \prod_{j=1}^l SU(M)_{i,j} \times \prod_{i=1}^k \prod_{j=1}^l 
SU(M)'_{i,j}
\eeq
The matter content is given by
\begin{center}
\begin{tabular}{ll}
{\bf Field} & {\bf Repr.} \\
$(A_1)_{i+1,j+1;i,j}$ & $(\fund_{i+1,j+1},\antifund\; '_{i,j})$ \\
$(A_2)_{i,j;i,j}$ & $(\fund_{i,j},\antifund\; '_{i,j})$ \\
$(B_1)_{i,j;i,j+1}$ & $(\fund\; '_{i,j},\antifund_{i,j+1})$ \\
$(B_2)_{i,j;i+1,j}$ & $(\fund\; '_{i,j},\antifund_{i+1,j})$
\end{tabular}
\end{center}
Observe that these field theories are generically chiral.
The superpotential is obtained by substitution of the surviving fields 
into the conifold superpotential. The outcome is the expression
\beqa
W= & \sum_{i,j} (A_1)_{i+1,j+1;i,j} (B_1)_{i,j;i,j+1} (A_2)_{i,j+1;i,j+1} 
(B_2)_{i,j+1;i+1,j+1} - \nonumber \\
& - \sum_{i,j} (A_1)_{i+1,j+1;i,j} (B_2)_{i,j;i+1,j} (A_2)_{i+1,j;i+1,j} 
(B_1)_{i+1,j;i+1,j+1}
\eeqa

The space (\ref{singukl}) has four curves of singularities  
radiating from the 
origin. Each is parametrized by either of the variables $x'$, $y'$, $z'$, 
$w'$, with the other variables equal to zero. It is possible to show that 
the Higgs branches corresponding to fractional D3 branes traveling along 
those curves actually exist in the field theory. 

There are other interesting branches. For instance, giving a vev to all 
the fields $(A_2)_{i,j;i,j}$, primed and unprimed factors with equal 
labels break to the diagonal combination. The final gauge group is 
$\prod_{i,j} SU(M)_{i,j}$. The surviving chiral multiplets are 
$(A_1)_{i,j;i-1,j-1}$, $(B_1)_{i,j;i,j+1}$, and $(B_2)_{i,j;i+1,j}$. The 
superpotential for these fields reads
\beq
W \propto \sum_{i,j} \left[ (A_1)_{i,j;i-1,j-1} (B_1)_{i-1,j-1;i-1,j} 
(B_2)_{i-1,j;i,j} \; - (A_1)_{i,j;i-1,j-1} (B_2)_{i-1,j-1;i,j-1} 
(B_1)_{i,j-1;i,j} \right]
\eeq
This field theory is that appearing on D3 branes on a 
$\IC^3/(\IZ_k\times \IZ_l)$ singularity \cite{dgm}, or equivalently 
\cite{hanur} on a $k\times l$ brane box model with trivial identifications 
of the unit cell \cite{hsu}. This fact will be recovered from the 
IIA brane configuration we are to propose as T-dual.

It is possible to construct a type IIA brane configuration such that the 
world-volume field theory on the D4 branes reproduces the theory just 
described. Consider the system of D4, NS and NS$'$ branes T dual to a set 
of D3 branes on the $\IZ_k$ quotient of the conifold, studied in 
Section~3. The models in this section are related to these theories by a 
further $\IZ_l$ quotient. The proposal is to also quotient the type IIA 
configuration by the twist
\beqa
x^4+ix^5 \to e^{2\pi i/l} (x^4+ix^5) \nonumber \\
x^8+ix^9 \to e^{2\pi i/l} (x^8+ix^9)
\label{quotlpt1}
\eeqa
which still preserves the $\NN=1$ supersymmetry \footnote{Brane  
configurations with orbifold quotients have been constructed in 
\cite{lpt1} (and in \cite{lpt2} with the inclusion of orientifold 
projections).}. This action is the 
counterpart of the quotient by $\beta$ in (\ref{quot2}). It must also be 
embedded on the Chan-Paton factors of the D4 branes. The final 
configuration provides a T dual realization of the field theory described 
above.

In order to check this claim, we can study the realization of some 
field theory Higgs branches in the branes setup. Obviously, some
mesonic Higgs branches correspond to splitting the D4 branes between pairs 
of identical NS (resp. NS$'$) branes, and moving them away along 45 (resp. 
89). Observe that this phenomenon, when described in the covering space of 
the orbifold, requires the D4 branes to move along with their mirror 
images. These branches correspond, in the singularity picture, to the 
motion of fractional D3 branes along two of the curves of singularities 
mentioned above.

To see one of the remaining mesonic Higgs branches in the brane configuration, 
let us split the D4 branes in the intervals into fractional 
branes with respect to the quotient (\ref{quotlpt1}). The fractional 
branes of different intervals can recombine to form a fractional D4 brane 
going around the whole $x^6$ circle. These objects can move away from the 
NS fivebranes along $x^7$, but not away from the orbifold singularity. 
Finally, there is a last mesonic branch whose realization in the brane 
picture is not manifest.

Let us turn to some baryonic Higgs branches. It is possible to remove 
{\em e.g.} the NS$'$ branes along $x^7$. If all such branes are removed, 
the remaining brane configuration contains NS branes along 012345, D4 
branes along 01236 and the $\IZ_l$ orbifold quotient acting on 4589. By 
performing a T duality along 6, the NS branes become an ALE space, {\em 
i.e.} a $\IZ_l$ orbifold quotient acting on 6789. The `superposition' of 
both quotients results in a local $\IC^3/(\IZ_k\times \IZ_l)$ singularity, 
at which D3 branes sit. Thus we recover the field theory Higgs branch we 
mentioned above.

It would be desirable to understand the meaning of possible 
transition in these more complicated singularities. Unfortunately, the 
IIA brane configuration we have constructed is not explicit enough to 
allow a direct comparison with the possible triangulations, as in the 
examples in section~3.3. We hope that a direct computation of the field 
theory using toric methods, along the lines of \cite{mp}, will help in 
clarifying this issue.

\medskip

\begin{figure}
\centering
\epsfxsize=3.5in   
\hspace*{0in}\vspace*{.2in}
\epsffile{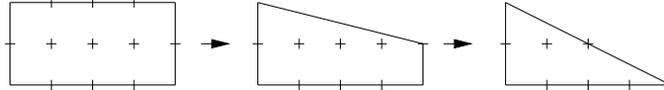}
\caption{\small The first toric diagram represents the singularity 
$x^2y^2=z^4w^4$. The singularity can be partially resolved to the 
$\IC^3/(\IZ_2\times \IZ_4)$ space. This process corresponds to the 
baryonic Higgs breaking mentioned in the main text.} 
\label{z2z4} 
\end{figure}

In analogy with Section~3  we can now consider models with different 
numbers of NS and NS$'$ branes. They provide T dual realization of the 
theory of D3 branes on singularities
\beq
x^l y^l=z^n w^m
\eeq
The superpotentials for these theories are obtained by starting with the 
orbifold model and following the Higgs breakings in the field theory. In 
Figure~\ref{z2z4} we illustrate how these breakings correspond to 
partial resolutions of the singular variety. The final model is reached 
from the initial one after a Higgssing along the baryonic branch mentioned 
above.

Even though the type IIA picture we have developed is somewhat 
complicated, it may prove useful to address certain issues. For instance, 
their lifting to M-theory seems to be quite tractable. However we will 
not pursue these points in the present work.

\subsection{Isolated singularities}

In this section we consider other quotients of the conifold variety, which 
generically lead to isolated singularities (in contrast with the previous 
cases, where the orbifold action had fixed complex planes). This is 
achieved by using discrete groups under which $x$, $y$, $z$ and $w$ are 
non-invariant. We will show a concrete example, which has a simple 
realization in terms of the underlying variables $a_i$, $b_i$. The  
generator $\theta$ of the group acts as
\beqa
\theta : & a_1 \to e^{2\pi i {1\over 2k}} a_1 \quad\quad\quad & x \to 
e^{2\pi i(r+1)/k} x \nonumber\\
& a_2 \to e^{-2\pi i{1\over 2k}} a_2 \quad\quad  \quad & y\to 
e^{-2\pi i(r+1)/k} y \nonumber \\
& b_1 \to e^{2\pi i{2r+1\over 2k}} b_1 \quad\quad  \quad & z\to 
e^{-2\pi i r/k} z \nonumber \\
& b_2 \to e^{-2\pi i{2r+1\over 2k}} b_2 \quad\quad  \quad & w\to 
e^{2\pi i r/k} w \nonumber 
\label{quotwist}
\eeqa

Notice that, if $p_1={\rm gcd(k,r)}\neq 1$, there are two complex planes 
(parametrized by $z$, $w$) of points fixed under a $\IZ_{p_1}$ action, 
generated by $\theta^{k/p_1}$. Analogously, if $p_2={\rm gcd}(k,r+1)\neq 
1$, there are two complex planes (parametrized by $x$, $y$) of points 
fixed under a $\IZ_{p_2}$ action, generated by $\theta^{k/p_2}$. For 
generic $k$, $r$, the singularity is isolated.

We choose to embed this action on the $SU(Mk)\times SU(Mk)'$ gauge degrees 
of freedom though the matrices (\ref{cp1}). The gauge group surviving the 
projection is $\prod_{i=1}^k SU(M)_i\times \prod_{i=1}^k SU(M)_i'$. The 
matter content is given by 
\begin{center}
\begin{tabular}{ll}
{\bf Field} & {\bf Repr.} \\
$(A_1)_{i+1,i}$ & $(\fund_{i+1},\antifund\; '_{i})$ \\
$(A_2)_{i,i}$ & $(\fund_{i},\antifund\; '_{i})$ \\
$(B_1)_{i+r,i}$ & $(\fund\; '_{i+r},\antifund_{i})$ \\
$(B_2)_{i,i+r+1}$ & $(\fund\; '_{i+r+1},\antifund_{i})$
\end{tabular}
\end{center}
The field theories arising from this construction are thus generically 
chiral. The superpotential is 
\beqa
W & = & \sum_{i=1}^k 
(A_1)_{i+1,i}(B_1)_{i,i-r}(A_2)_{i-r,i-r}(B_2)_{i-r,i+1} 
- \nonumber \\
& & \sum_{i=1}^k  (A_1)_{i+1,i} (B_2)_{i,i+r+1} (A_2)_{i+r+1,i+r+1} 
(B_1)_{i+r+1,i+1}.
\eeqa

For generic values of $k$, $r$ there are no mesonic branches, 
corresponding to the fact that there are no curves of singularities 
emanating from the origin. However, there is an interesting baryonic 
branch, which is obtained by giving diagonal vevs to, say, all the 
$(A_2)_{i,i}$ fields. The superpotential interaction for the remaining 
fields is
\beqa
W\; = \; (A_1)_{i+1,i} (B_1)_{i,i-r} (B_2)_{i-r,i+1} \; -\;
(A_1)_{i+1,i} (B_2)_{i,i+r+1} (B_1)_{i+r+1,i+1}
\eeqa

This field theory can also be obtained using a brane box model where the 
$k\times 1$ box unit cell has its vertical sides identified trivially, and 
its horizontal sides are identified up to a shift of $r$ boxes \cite{hsu}. 
Equivalently \cite{hanur}, this theory arises from D3 branes on a 
$\IC^3/\IZ_k$ singularity, where the $\IZ_k$ acts by multiplying the three 
complex planes by $e^{-2\pi i/k}$, $e^{-2\pi i r/k}$, $e^{2\pi i(r+1)/k}$, 
respectively.

In principle, it would be possible to construct a type IIA brane 
configuration reproducing this field theory. The most systematic way would 
be to embed the action (\ref{quotwist}) in a $\IZ_n\times \IZ_m$ group of 
the type studied in section~5.1, in such a way that the subgroup can be 
recovered by quotienting by a suitable equivalence relation. The `parent' 
singularity has a T dual IIA brane configuration which has been already 
described. The T brane configuration of the singularity of interest is 
obtained upon making the appropriate identifications in the `parent' brane 
model. In particular these will amount to identifying fractional 
branes of different intervals. However, since this representation is 
rather involved, and its usefulness is doubtful, we will not extend on a 
general recipe.

\section{Introduction of orientifold planes}

The inclusion of orientifold planes in the type IIA brane picture is 
straightforward. For simplicity, we will only consider NS fivebranes at 
right angles. Even in this restricted context, there exist different 
patterns of orientifolded theories. We now turn to sketching them.

Starting with a brane configuration of the type studied in Section~3, we 
can include an orientifold four-plane, parallel to the D4 
branes. These are $\NN=1$ versions of the models studied in \cite{lll}.
Recall that the charge 
of this orientifold changes sign whenever it crosses a NS fivebrane 
\cite{johnson}. The general gauge group has a structure $Sp$$\times 
SO$$\times Sp$$ \ldots \times SO$. Consistency requires the total number 
of 
NS fivebranes to be even. Once a distribution of NS and NS$'$ branes is 
chosen, the field theory content and superpotential can be read off from 
those of the non-orientifolded theory by simply projecting onto invariant 
fields. The effect of this projection on the gauge group has been already 
mentioned; also, after the projection, strings stretching between D4 
branes of adjacent intervals give rise to only one chiral multiplet 
in the corresponding bi-fundamental representation. The surviving fields 
have a superpotential inherited from the non-orientifolded case.

Instead, we can introduce orientifold six-planes (O6 planes), extending 
along 0123789 \footnote{The configurations with O6$'$ planes are 
obtained from those with O6 planes by exchange of the coordinates 45 and 
89, and so do not provide new models.}. These are $\NN=1$ versions 
of the models studied in 
\cite{uranga}. The orientifold projection inverts the $x^6$ coordinate, 
and so, working on the double cover, the distribution of NS fivebranes on 
the circle must be $\IZ_2$ 
symmetric. There are two O6 planes, sitting at opposite position in the 
$x^6$ circle. The general pattern for the field theory gauge group is
\beq
G_1 \times SU \times \ldots \times SU \times G_2
\label{sausage}
\eeq
The $x^6$ intervals corresponding to the middle $SU$ factors are mapped to 
some `mirror' intervals, and thus this sector of the theory is 
unrestricted, and has the structure of a non-orientifolded theory, namely 
the group factors are $SU$, and strings stretched between D4 branes yield 
two chiral fields, in the bi-fundamental and conjugate representations. 
The superpotential is also found using the rules in Section~3.

The `end' sectors in the chain (\ref{sausage}) correspond to the 
$x^6$ intervals closest to the O6 planes. There are several possibilities 
for this sector, depending on the distribution of fivebranes near the O6 
plane. The basic building blocks have appeared in \cite{lust} 
\footnote{The $\NN=2$ building blocks were introduced in \cite{ll}.},  and  
we schematically give them in the following table
{\small
\begin{center}
\begin{tabular}{ccccc}
{\bf Config.} & {\bf Group} & {\bf Matter} & {\bf Superpot.} &{\bf 
Comments} \\
$NS-O6^{+}-NS$  &  $SO$     &   $S$        & $FS{\tilde F}$ & $N=2$\\
$NS-O6^{-}-NS$  &  $Sp$     &   $A$        & $FA{\tilde F}$ & $N=2$\\
$NS'-O6^{+}-NS'$  &  $SO$     &   $A$        & $FA{\tilde F}$ &  \\
$NS'-O6^{-}-NS'$  &  $Sp$     &   $S$        & $FS{\tilde F}$ \\
$NS-(NS,O6^{+})-NS$  &  $SU$     &   $S,{\tilde S},X$ & $SX{\tilde 
S}-FX{\tilde F}$ & $N=2$\\
$NS-(NS,O6^{-})-NS$  &  $SU$     &   $A,{\tilde A},X$ & $AX{\tilde 
A}-FX{\tilde F}$ &  $N=2$\\
$NS'-(NS,O6^{+})-NS'$ & $SU$     &   $S,{\tilde S}$ & $S{\tilde 
S}F{\tilde F}$ & \\
$NS'-(NS,O6^{-})-NS'$  &  $SU$     &   $A,{\tilde A}$ & $A{\tilde 
A}F{\tilde F}$ &\\
$NS-(NS',O6)-NS$  &  $SU$     &   $A,{\tilde S}, 8T$ & ${\tilde S}TT + 
F{\tilde F}A {\tilde S}$ &  Chiral\\
$NS'-(NS',O6)-NS'$  & $SU$     &   $A,{\tilde S},8T,X$ & ${\tilde S}TT  
- A{\tilde S}X + FX{\tilde F}$ &  Chiral\\
\end{tabular}
\end{center}
}

The notation is: $S$($A$) denote two-index symmetric (antisymmetric) 
representations, whereas $X$ and $T$ denote the adjoint and fundamental 
representations of $SU$, respectively. The field $F$ is a bi-fundamental 
field arising at the outer NS fivebranes in this sector. Tildes denote 
conjugate representations. When a NS fivebrane is stuck at the O6 plane, 
we write them both inside parentheses, otherwise the sequential order 
reproduces the ordering along $x^6$. 

All these sectors can be introduced as endpoints in the chain 
(\ref{sausage}). Once a given choice is made, it is straightforward to 
write down the field theory content and interactions.

A last type of configuration involving only NS fivebranes at right angles 
is possible by using NS fivebranes rotated $\pm 45^0$ in the 45, 89 
direction. We 
denote these as NS1 and NS2 branes, respectively. Notice that the 
orientifold plane maps NS1 to NS2 branes, since it inverts the coordinates 
45, while leaving 89 invariant. Thus, these type of branes cannot be stuck 
at the O6 planes. Again, the general gauge group has the structure 
(\ref{sausage}), where the middle sector is that of a 
non-orientifolded theory. The endpoint sectors consist on a group factor 
$SO$ or $Sp$ with no matter, save for bi-fundamentals. The superpotential for 
these last is quartic. If one allows for branes at arbitrary angles, it is 
possible to interpolate between this case and the one analyzed above.

It is clear that a large variety of models can be obtained if we also 
allow to quotient by $\IZ_k$ actions in these brane configurations, in 
analogy with the models studied in Section~5. Moreover, it is also 
possible to include some orientifold projection acting along with this 
orbifold action. The scenarios can be complicated at wish, and it would be 
interesting to see what they can teach about orientifolding conifold (and 
generalized conifold) geometries in the T dual side. We leave all these 
points for future research.

\section{Final Comments}

We have proposed a T duality between D3 branes on some non-orbifold 
spaces and type IIA configurations of relatively rotated NS fivebranes and 
D4 branes. We expect a rich interplay between both realizations of the 
same field theories. Eventually, it may lead to an AdS realization of the 
diverse phenomena observed in lifting the IIA brane configuration to 
M-theory.

It would be desirable to have a better understanding on how the geometry 
of the singularity determines the field theory. We hope the examples we 
have developed allow to infer some rules in this respect. In particular, our 
correspondence between different triangulations of the toric diagram with 
different field theories points in this direction, and is worth of further 
exploration. These rules may eventually provide the key to understanding
more general singularities. At this point, we 
should stress that we have restricted to brane configurations with 
fivebranes at right angles, and that relaxing this constraint leads to a 
larger family of T-dual singularities. 

\bigskip

\centerline{\bf Acknowledgements}

I am pleased to thank A.~Klemm, and M.J.~Strassler for useful 
discussions, and especially to K.~Dasgupta for discussion and for 
sharing with me his related results with Sunil Mukhi. I am also grateful 
to M.~Gonz\'alez for advice and support.
This work is supported by the Ram\'on Areces Foundation (Spain).

\bibliography{conifold}
\bibliographystyle{utphys}
\end{document}